%% file: big_data_main.tex
\documentclass[conference]{IEEEtran}
\IEEEoverridecommandlockouts

\usepackage{amsmath,amssymb,amsfonts}
\usepackage{graphicx}
\usepackage{textcomp}
\usepackage{xcolor}

\usepackage{url}
\usepackage{subcaption}

\usepackage[noend]{algpseudocode}
\usepackage{algorithm}
\usepackage{bbm}

\usepackage{cite}

\def\BibTeX{{\rm B\kern-.05em{\sc i\kern-.025em b}\kern-.08em
    T\kern-.1667em\lower.7ex\hbox{E}\kern-.125emX}}
\begin{document}

\title{Trustworthy Scheduling for Big Data Applications}

\author{\IEEEauthorblockN{Dimitrios Tomaras, Vana Kalogeraki}
\IEEEauthorblockA{Dept. of Informatics\\
\textit{Athens University of Economics and Business}\\
{tomaras,vana}@aueb.gr}
\and
\IEEEauthorblockN{Dimitrios Gunopulos}
\IEEEauthorblockA{Dept. of Informatics and Telecommunications\\
\textit{National and Kapodistrian University of Athens}\\
dg@di.uoa.gr}
}
\maketitle

\begin{abstract}
Recent advances in modern containerized execution environments have resulted in substantial benefits in terms of elasticity and more efficient utilization of computing resources.
Although existing schedulers strive to optimize performance metrics like task execution times and resource utilization, they provide limited transparency into their decision-making processes or the specific {\it actions} developers must take to meet Service Level Objectives (SLOs). 
In this work, we propose X-Sched, a middleware that uses explainability techniques to generate actionable guidance on resource configurations that makes task execution in containerized environments feasible, under resource and time constraints. 
X-Sched addresses this gap by integrating counterfactual explanations with advanced machine learning models, such as Random Forests, to efficiently identify optimal configurations. This approach not only ensures that tasks are executed in line with performance goals but also gives users clear, actionable insights into the rationale behind scheduling decisions.
Our experimental results validated with data from real-world execution environments, illustrate the efficiency, benefits and practicality of our approach.
\end{abstract}


\input{intro}

\input{model}

\input{approach}

\input{evaluation}
\input{related}

\input{conclusion}

\section*{Acknowledgment}

This research has been supported by the European Union through the Horizon Europe AutoFair project (No. 101070568) and the Horizon Europe CoDiet project (No. 101084642).

\bibliographystyle{IEEEtran}
\bibliography{refs}

\end{document}

%% file: intro.tex
\section{Introduction}

In today’s increasingly complex big data application landscape, task automation frameworks such as MLflow, Airflow, and Kubeflow have emerged to streamline development and execution in containerized cloud and edge environments. 
Despite the increasing sophistication of cloud platforms, the critical task of \textit{resource sizing} is still left to practitioners, who must determine the appropriate container sizes and instance configurations. This decision is vital, as selecting optimal resource allocations can lead to faster execution times at a lower cost.
Moreover, there is no rule-of-thumb, resource configuration impacts vary widely across applications.
This makes predicting task execution times particularly challenging, often requiring historical data from similarly sized applications. 
As a result, practitioners may often request more resources than needed (\textit{over-provisioning}) or very few (\textit{under-provisioning})\cite{yadav2021resource} or choose a \textit{default} recommendation from the cloud provider, which, typically, is not the most appropriate for their applications.

Prior work in the literature \cite{DBLP:conf/bigdataconf/Zhang0WSL19,hu2019spear,DBLP:journals/cluster/ChengHTSCL22,DBLP:conf/sosp/NarayananKAKAKB21,DBLP:conf/sc/WangW00020,DBLP:conf/bigdataconf/LiuLMB23,drougas} has underscored the significance of \textit{resource sensitivity}\cite{DBLP:journals/jss/TurinBDDJT23}, 
exploiting resource parameters, such as memory, CPU or input data size to accurately predict task completion times. A plethora of techniques 
that aim to optimize task completion times, SLO targets or request service rates have been proposed. They can be broadly categorized into two main methodologies: (i) determining the optimal configuration by solving an Integer Linear Programming (ILP) problem, using exploration processes like GridSearch algorithms\cite{DBLP:conf/sc/WangW00020}, Model Agnostic Meta Learning\cite{DBLP:conf/cloud/WangYYWLSHZ21}, sequential model-based optimization (SMBO)\cite{DBLP:conf/nips/SnoekLA12}, 
or (ii) leveraging historical execution runs to predict task completion times and generate resource configuration policies applying Bayesian Optimization or Deep Reinforcement Learning (DRL) techniques\cite{hu2019spear,DBLP:journals/cluster/ChengHTSCL22,DBLP:conf/bigdataconf/Zhang0WSL19,funika2023automated}.

Nevertheless, the key limitation of all these approaches lies in their lack of guidance, which renders them black-box methods, obscuring their decision making process and preventing them from providing explanations or offering alternative configuration suggestions for setting up the task execution environment.
As a result, users are not guided on choosing a suitable configuration of container resources for their tasks prior to execution, leaving them to rely on trial-and-error approaches. For instance, the AWS Compute Optimizer chooses not to provide recommendations for all the EC2 instance families; instead, it restricts its analysis to a specific historical time-window and relies on the CloudWatch Agent installation and configuration. Furthermore, these approaches have shown to impose a significant burden on cluster administrators, especially for unknown workloads 
\cite{DBLP:conf/middleware/EismannBGAHK21}, which are quite limited as they cannot be run on user tasks. 
Thus, developing a system that actively assists users in selecting container resources, especially when critical activities fail to be scheduled due to imposed constraints, becomes crucial.

{\bf Why developing an explainable scheduler?} 
The notion of \textit{explainability} has recently gained attention as a promising way to make black-box ML and AI models more understandable, enabling end-users to grasp the reasoning behind their decisions, address biases and resolve potential conflicts. 
Practitioners are very likely to trust systems if they feel they have more control on the decision making, the system behavior meets their expectations and they perceive the system's decisions being transparent and interpretable. 
Although black-box schedulers, such as SLURM\cite{DBLP:conf/hipc/ChadhaJG20}, have been successfully deployed for decades, their success has relied on relatively stable environments and rule-based heuristics whose behaviors, while opaque, are predictable. The increasing heterogeneity of big data workloads and the demand for fairness and accountability all necessitate a paradigm shift towards explainable scheduling. Explainability is therefore not merely a desirable feature but an emerging requirement for sustaining trust and transparency in modern big data processing frameworks. 
Recent studies %\cite{baruah2023towards,DBLP:conf/aaai/CyrasLMT19} 
\cite{DBLP:conf/aaai/CyrasLMT19} have highlighted \textit{schedulability explanations} as a viable solution, empowering users to gain a deeper understanding of the system and predict its behavior more accurately. 
This work emphasizes the need to equip users with actionable guidance for adjusting container configurations under deadline constraints. To address this challenge, we advocate for 
a middleware with built-in explainability, offering transparent and interpretable decisions that extend beyond container resource allocation to broader applications in containerized systems (e.g., proposing alternative deadlines or optimizing execution costs).

Determining appropriate tasks' resource sizing 
to satisfy service level objectives  
is a challenging problem. 
Several factors including 
the variability in task workloads (dynamic requests fluctuating over time) as well as inter-dependencies among tasks 
({\it i.e.,} applications composed of task pipelines or direct acyclic graphs) 
make the selection of suitable container resources challenging. 
The problem of identifying the appropriate configuration for a given task deadline may 
require exploring all possible alternative predictions, which is both time-consuming and cost-inefficient\cite{ali2023hyperparameter}.
Finally, generating explanations (i.e providing users with sets of actions to properly configure container resources) requires an understanding of how each task, and subsequently its completion time, is affected by resource allocation. However, this is a tedious process as each task is treated as a black box, making a straightforward approach unavailable.

\begin{figure}[!t]
\begin{minipage}{0.475\linewidth}\centering
\centering
\includegraphics[width=\textwidth,height=1.1in]{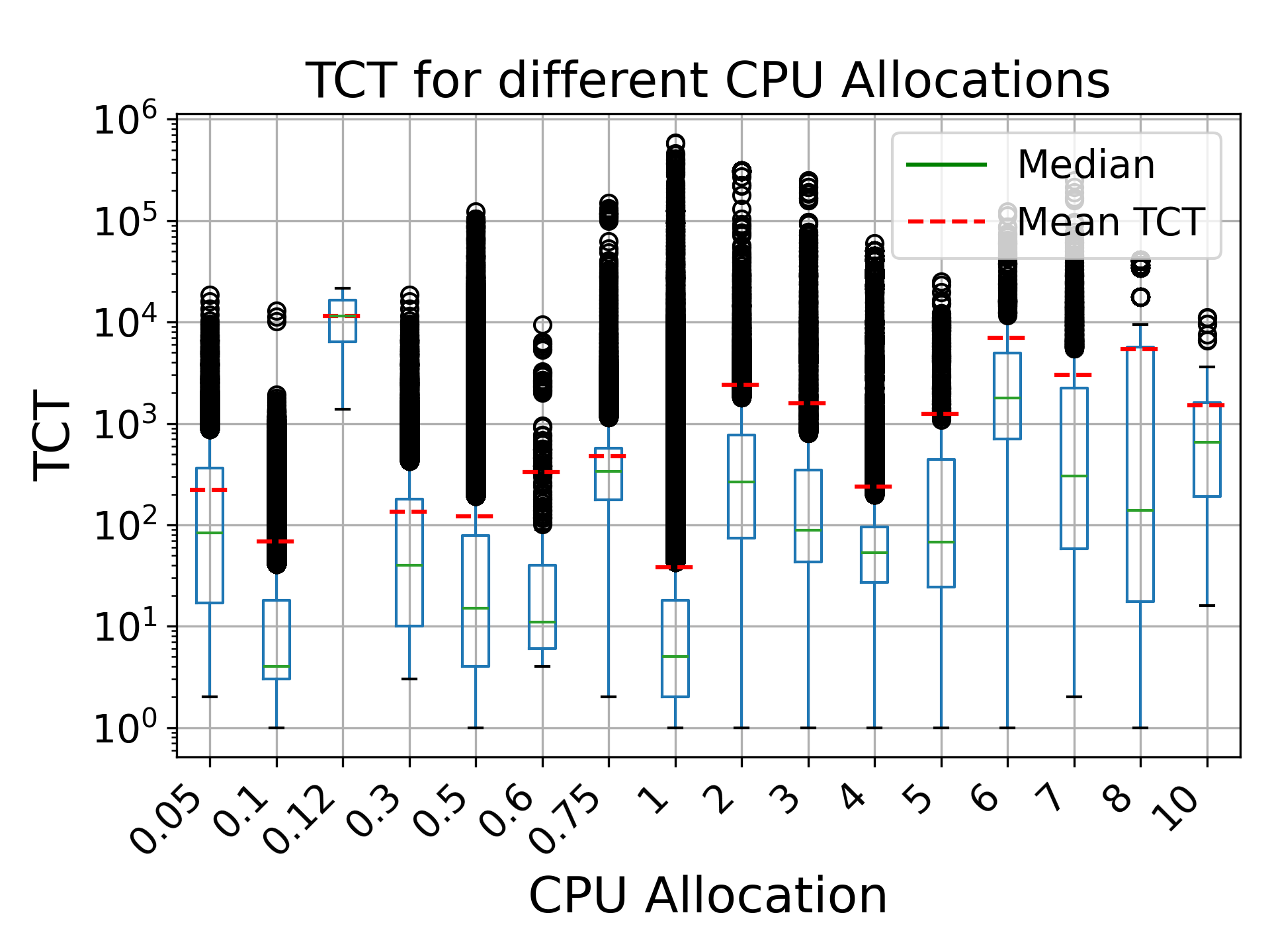}
\caption{Task completion time (TCT) per CPU allocation}
\label{fig:tct_cpu_mean}
\end{minipage}\hfill
\begin{minipage}{0.475\linewidth}\centering
\centering
\includegraphics[width=\textwidth,height=1.1in]{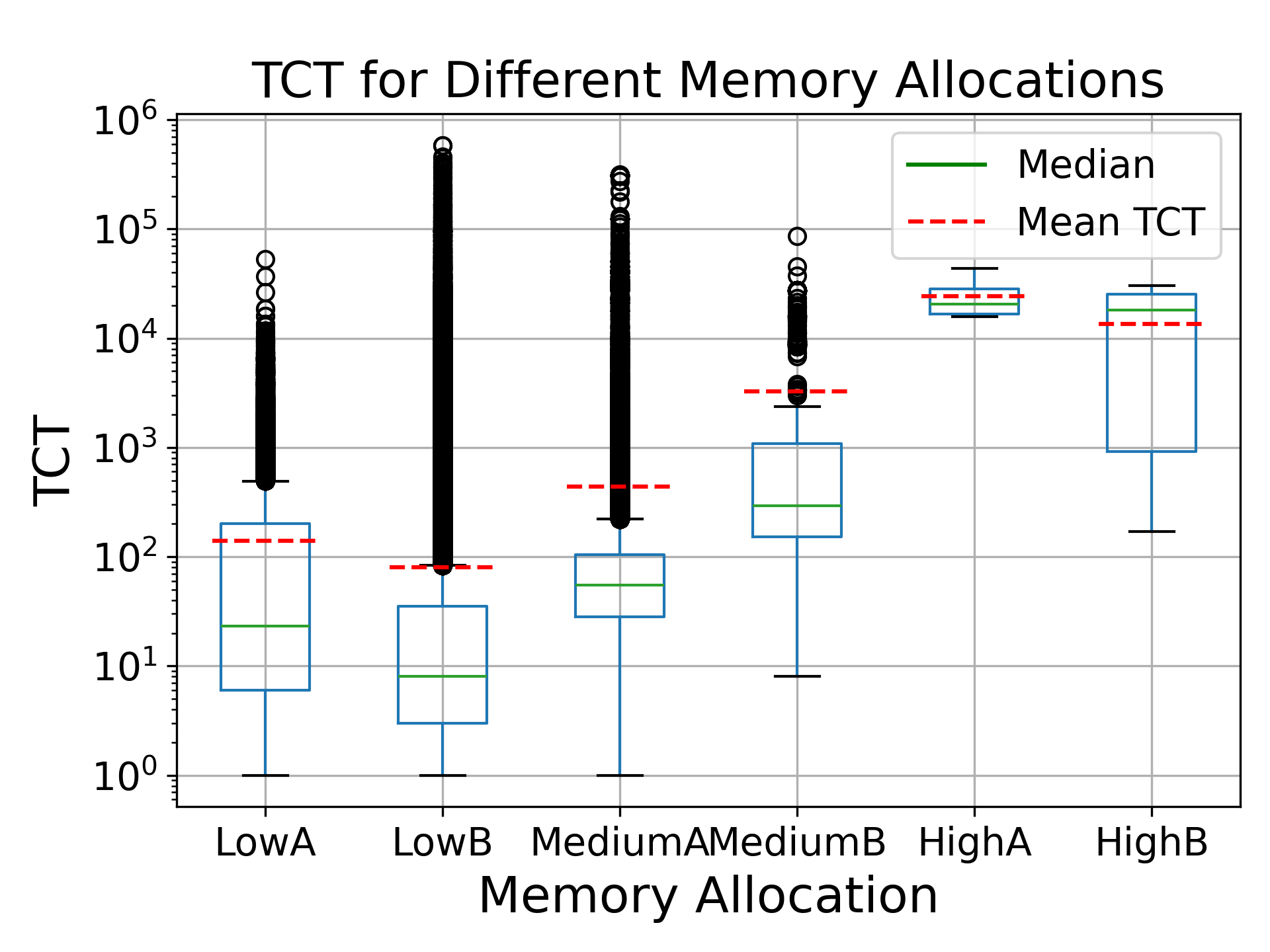}
\caption{Task completion time (TCT) per memory allocation}
\label{fig:tct_mem_mean}
\end{minipage}\hfill
\end{figure}

{\bf Motivating example:} To further motivate our work we delve into a dataset of tasks from Alibaba's production cluster\cite{alibabadata}
that provides detailed statistics for co-located workloads of long-running and batch jobs over a course of 24 hours. Jobs typically comprise several tasks and each task has a number of instances.
For each task they report the task completion time 
and the corresponding container configuration (CPU, memory), these are selected from a predefined set of memory and CPU configurations (the CPU cores are divided into 16 resource allocation categories while the memory sizes correspond to 6 allocations). 
In Figures \ref{fig:tct_cpu_mean} and \ref{fig:tct_mem_mean} we illustrate the task completion time as a function of the predefined CPU and memory allocations respectively. As Figure \ref{fig:tct_cpu_mean} shows, 
the number of CPU cores is not strongly correlated to the task completion times, and in fact the task completion times rise slightly when the number of CPU cores increases. The reason for this counterintuitive result is that the tasks are not identical, and the user attempts to get the right CPU provisioning of each task. As the user would typically provision more CPUs for the more computationally intensive tasks, the figure shows that the task completion time tends to remain relatively stable despite the number of CPU cores used. 
Figure \ref{fig:tct_mem_mean} shows the corresponding effect on memory allocation. Our first observation is, that, users make heuristic attempts to properly provision the different tasks. The second important observation is that these heuristic attempts are not very good: the large variance of the task completion times strongly indicates that there is often a large difference between what the user expects to be adequate provisioning in CPU cores or memory and what is actually required.

Clearly, 
there is no rule-of-thumb or straightforward guideline. 
Our goal in this work is to build a middleware that can overcome the gap created by black-box models 
offering a system that is not only capable of meeting the scheduling requirements of containerized tasks, but most importantly, delivers transparent, explainable insights into container resource provisioning decisions.

{\bf Contributions. } We summarize our contributions as follows:
\begin{itemize}
    \item We propose an explainable middleware, X-Sched, that uses explainability techniques to generate actionable guidance on resource configurations that makes task execution in containerized environments feasible under resource and time constraints. 
    The X-Sched Runtime aims to generate 
    configuration changes to help users
    make smart decisions, extending beyond a classical container allocation problem.
    \item We developed the X-Sched Library to allow users to define and support a wide variety of tasks, ranging from simple to more complex tasks (e.g. ML models).
    \item Our detailed experiments with respect to feasibility and density of solutions, illustrate its practicality and wide applicability in real-world scenarios.
\end{itemize}

%% file: model.tex
\section{System Model \& Background }

\subsection{Notation}

We consider a platform that can run big data tasks\cite{DBLP:conf/acsos/PeriTK23,DBLP:conf/cloud/JoosenHASDWB23}. Typically, tasks run in container images in a distributed fashion over a container orchestrator such as Kubernetes, Docker Swarm  or Apache Mesos.  
Let us denote as $\mathcal{S}$ the containerized platform,  
the $k$ types of tasks as $t_{k}$ and the task's container as $\mathcal{C}_{t_{k}}$. 
The users typically specify the size of the container, in terms of the amount of memory
$\mathcal{C}_{t_{k}}^{mem}$ (e.g. AWS's m8g.medium provides 4GB of memory compared to 8GB for m8g.large), number of CPU cores $\mathcal{C}_{t_{k}}^{cpu}$ (e.g. m8g.4xlarge provides 12 vCPUs while m8g.12xlarge has 48 vCPUs) and number of replicas 
$\mathcal{C}_{t_{k}}^{rep}$ 
for tasks that serve 
simultaneous requests. Commercial systems may also provide predefined category groups when selecting these features.
For instance, in the Alibaba production clusters\cite{alibabadata}, memory size is normalized into {\it low}, {\it medium} or {\it high} memory. 
We assume exclusive per-container resource configuration for each task, without any resource sharing. Tasks are also associated 
with timing constraints in the form of Service Level Objectives (SLOs) and have the same priority. We formally model a task as follows:
\begin{align}
    t_{k} = \{id_{t_k},\mathcal{C}_{t_{k}}^{mem},\mathcal{C}_{t_{k}}^{cpu},\mathcal{C}_{t_{k}}^{rep},T_{t_{k}}^{start},T_{t_{k}},t_{k}^{dl},st_{t_{k}},tp_{t_{k}}\}
\end{align}
\noindent Each task $t_{k}$ has a start time $T_{t_{k}}^{start}$, an end time $T_{t_{k}}^{end}$, 
a \textit{deadline} $t_{k}^{dl}$ within which its execution must complete and is associated with the amount of time $T_{t_{k}}=T_{t_{k}}^{end}-T_{t_{k}}^{start}$ it takes to execute. The task completion time depends not only on the complexity of the task's algorithmic code, but more importantly depends on the hosting environment ({\it i.e.,} memory, CPU cores, number of replicas) of the container in which task $t_{k}$ executes\cite{DBLP:conf/middleware/VelpRS20,DBLP:journals/tsc/LouLTJZ23}. A task also has a unique id $id_{t_k}$, a status $st_{t_{k}}$, {\it i.e.,} "running" or "finished", and also a type 
$tp_{t_{k}}$ which corresponds to the type of the task (\textit{lra} denotes long-running applications while \textit{bpa} denotes batch processing applications).

\subsection{Explainability Background}

In our work we leverage the theory of explainability to help users gain deeper insights into system behavior and more accurately predict the schedulability of their application tasks.

\noindent{\bf Definition: (Explainability)} Explainability in machine learning is defined as the property of a model or a system that enables the user to understand the contribution and influence of individual input features on the model's prediction. More formally, assume a machine learning model, such as a classifier $f$, defined as $f(\textbf{X}):\mathbb{R}^{n}\rightarrow \mathbb{R}$, 
where $\textbf{X}=(x_{1},x_{2},..,x_{n})$ represents the vector of input features. Explainability aims to provide a mathematical framework to quantify the impact of each feature $x_{n}$ on the output $f(\textbf{X})$. Explainability is often approached by decomposing the prediction $f(\textbf{X})$ into contributions from each feature $x_{n}$, $f(\textbf{X}) = \phi_{0}+\sum_{m=1}^{n}\phi_{m}$, 
where $\phi_{0}$ represents the baseline or average model output when no features are present and $\phi_{m}$ represents the contribution of feature $x_{n}$ to the prediction.

We opt for the method of counterfactual explanations\cite{wachter2017counterfactual,DBLP:conf/fat/MothilalST20,pmlr-v202-ley23a} in order to identify the minimal changes required to alter the prediction of a classifier. Counterfactual explanations identify the aspects of the input data that most influence the model's decisions, offering clear insights into how small changes in inputs could lead to different outcomes.
We first develop an appropriate classification function that predicts the schedulability of a task for a given deadline, 
and then, using the counterfactual explanations, we identify the minimal changes (if needed) so that the task becomes schedulable.

%% file: approach.tex
\section{The X-Sched Library}

We developed the X-Sched Library to allow users to
define and support a wide variety of tasks, ranging from simple to more complex tasks (e.g. ML models).
The X-Sched library offers an abstraction, i.e. a scheduling entity class, namely \textit{XSchedTask}, that is utilized to define tasks, ranging from single stand-alone tasks (e.g. a matrix multiplication function) to more complex tasks (e.g. a Random Forest model)\cite{rogowski2022mpi4py}. XSchedTask, as a scheduling entity, encapsulates all the necessary information of the task, i.e., its id, status, type, initial container configuration, task deadline, etc. It also serves as the interface for the definition of complex tasks, such as the training of ML models (e.g. a DNN model). 
Defining complex tasks introduces the additional requirement of designing and implementing communication and synchronization mechanisms. 
Yet, existing task automation frameworks lack built-in support for exchanging data, code, and model artifacts, capabilities essential for modularity, flexibility, and interoperability with other systems. 
At the same time, libraries that implement a subset of these mechanisms like \textit{mpi4py} have several limitations: they are tightly coupled with MPI backends, are process-based, usually require an external launcher, lack support for HTTP or cloud-native messaging layers like gRPC, as well as, native hooks to async communication layers.

For all these aforementioned reasons, we have implemented and provide these primitives\cite{mai2020kungfu,zhang2019mllib,kehrer2019serverless} in X-Sched library 
to cover a wide variety of tasks: (a) $Broadcast$, for synchronous training and parameter broadcasting tasks, (b) $AllGather$, for federated learning tasks, (c) $AllReduce$, for gradient synchronization (a typical task in DNN training) and parallel model evaluation tasks, (d) $Gather$, for data preprocessing and distributed Extract-Transform-Load pipelined tasks, (e) $Reduce$, for tasks that require a centralized aggregation e.g. job reporting, and (f) $Scatter$, for data-parallel training tasks. For the above communication schemes, we also implemented their synchronized and async versions using protocols like the Bulk Synchronous Parallel (BSP) and the Asynchronous Parallel (ASP)\cite{jiang2021towards,jiang2022fedmp}. 
The ASP protocol is more appropriate for best effort tasks for which we may need only a single or a set of the fastest answers, whereas the BSP protocol is more appropriate for tasks requiring synchronization. For instance, we may either have tasks, for which, just a fast response is adequate (best effort tasks) or we may need to run tasks that require synchronization before propagating the results to other tasks (like the training Random Forests with a number of parallel decision trees).
The X-Sched library implements the XSchedTask and offers all these primitives. X-Sched is integrated with MLflow and is built upon MLflow's \textit{mlflow.pyfunc} and \textit{MLflow plugins}.

\begin{figure}[!t]\hfill
\begin{minipage}{\linewidth}\centering
\centering
\includegraphics[width=\textwidth]{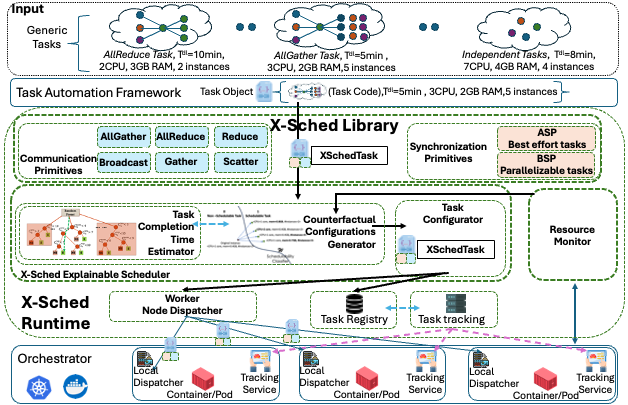}
\caption{X-Sched middleware}
\label{fig:middleware}
\end{minipage}\hfill
\end{figure}

\section{The X-Sched Runtime}

X-Sched provides inherently interpretable mechanisms that not only address the definition of containerized tasks but, most importantly, deliver valuable insights into container resource provisioning. The X-Sched Runtime's primary goal (shown in Figure \ref{fig:middleware}) is, given a XSchedTask instance $\textbf{x}_{t_{k}}$,
to derive a set of possible configuration solutions, {\it i.e.,} in the form of actionable counterfactual explanations, along with the specific actions that yield each solution.

\subsection{Building Schedulable XSchedTasks}

{\bf Actions. } 
Similar to related works\cite{wachter2017counterfactual,DBLP:conf/fat/MothilalST20,pmlr-v202-ley23a}, we denote as $\mathbb{A}$ the set of all possible actions (which is potentially infinite), where an \textbf{action} $a \in \mathbb{A}$ is a set of changes to feature values, e.g., $a = \{\text{CPU} \rightarrow +2\text{ cores}, \text{instance-num} \rightarrow +2\}$, which, when applied to an instance $\textbf{x}_{t_{k}}$, results in a \textbf{counterfactual} instance $\textbf{x}_{t_{k}}' = a(\textbf{x}_{t_{k}})$. 
The scheduler generates a set of $q$ counterfactual explanations, represented as container configuration tuples $\{cf_{1},cf_{2},...,cf_{q}\}$. Each tuple corresponds to a configuration where the task is predicted to meet its deadline requirement. 
To achieve that, the literature suggests two main approaches: i) generate actions that change the feature values of a \textit{single individual instance}, commonly known as \textit{local counterfactual explanations} \cite{DBLP:conf/fat/MothilalST20}, or ii) generate actions that provide directions for actions for \textit{more than one individual instance}, known as global counterfactual explanations (e.g. using frameworks like GLOB-CE\cite{pmlr-v202-ley23a}). 
However, the latter one is not applicable in our setting, since we focus on individual tasks.

{\bf Schedulability Boundary Function.} 
To generate the set of $q$ counterfactual explanations, $\{cf_{1},cf_{2},...,cf_{q}\}$, we use a classifier function $f=\mathcal{W}(\textbf{x}_{t_{k}})$ that evaluates whether one of the counterfactual explanations makes a task $\textbf{x}_{t_{k}}$ to become schedulable. 
The output $\mathcal{Y}$ of the function $\mathcal{W}(\textbf{x}_{t_{k}})$ is equal to 1 if the task is schedulable ($\mathcal{Y}=1,\mathcal{W}(\textbf{x}_{t_{k}})=1$) or 0 ($\mathcal{Y}=0,\mathcal{W}(\textbf{x}_{t_{k}})=0$) if it is non-schedulable.

The goal of the scheduling algorithm is, for a given task $t_{k}$, to derive actions that correspond to (memory, CPU cores and instances) tuples $\langle \mathcal{C}_{t_{k}}^{mem},\mathcal{C}_{t_{k}}^{cpu},\mathcal{C}_{t_{k}}^{rep}\rangle$ 
so that it completes its execution within the user deadline. 
For this purpose, we need to model the task's $t_{k}$ completion time by a function $\mathcal{G}_{t_{k}}(\bullet)$, $\mathcal{G}_{t_{k}}: \mathbb{R}^{n}\rightarrow \mathbb{R}$, which takes as input the container size and outputs the estimated amount of time for the task to complete given these container's resources, as:
\begin{align}
    T_{t_{k}}=T_{t_{k}}^{end}-T_{t_{k}}^{start}=\mathcal{G}_{t_{k}}(\mathcal{C}_{t_{k}}^{mem},\mathcal{C}_{t_{k}}^{cpu},\mathcal{C}_{t_{k}}^{rep})
\end{align}

We refer to the classifier function as the \textit{Schedulability Boundary Function}, which enables the estimation of the schedulability boundary for a task, allowing it to be classified as either schedulable or non-schedulable based on a given deadline.
This function classifies the difference $t_{k}^{dl}-\mathcal{G}_{t_{k}}(\mathcal{C}_{t_{k}}^{mem},\mathcal{C}_{t_{k}}^{cpu},\mathcal{C}_{t_{k}}^{rep})$ 
between the given task deadline and the task completion time as either one (indicating the deadline is met) or zero (indicating the deadline is missed). More formally,
\begin{align}
      \mathcal{W}(\textbf{x}_{t_{k}})=\begin{cases}
        1,  & \text{if }  t_{k}^{dl}-T_{t_{k}}\geq 0\\
        0,  & \text{if }  t_{k}^{dl}-T_{t_{k}} < 0
    \end{cases}
\end{align}

Our goal is to determine the Schedulability Boundary Function effectively, as the value of the function $\mathcal{G}_{t_{k}}(\mathcal{C}_{t_{k}}^{mem},\mathcal{C}_{t_{k}}^{cpu},\mathcal{C}_{t_{k}}^{rep})$ for a given container configuration should be estimated from the set of historical task data (as discussed in Section \ref{ssec:trees}). 
Thus, we define a Schedulability Boundary Function loss metric, $loss(\mathcal{W}(cf_{u}),y_{cf})=||\mathcal{W}(cf_{q})-y_{cf}||_{l_2}$, that minimizes the $l_2$-distance between the $\mathcal{W}(cf_{q})$ prediction for the counterfactual explanation $cf_{q}$ and the desired outcome $y_{cf}$ (which in our case is equal to $\mathcal{Y}=1$ for making the task schedulable).

{\bf Counterfactual Generation Constraints.} 
While counterfactual explanations enhance the likelihood that at least one of the generated examples will be actionable for the user, they may result in altering a broad range of features or propose solutions that are distant from the original user request. 
Our explainable scheduler is designed to generate solutions that are (a) 
in \textit{close proximity} to the original instance, (b) \textit{diverse}, (c) 
encompass \textit{changes to a small amount of features} and are (d) 
\textit{feasible}.

{\bf Optimization problem.} Given the above, 
we define a combined loss function $\mathcal{L}$ for all the generated counterfactuals of the original instance $\textbf{x}_{t_{k}}$, and our objective is to minimize this loss function:
{
\begin{align}\small
    \mathcal{L}(\textbf{x}_{t_{k}})=\arg\min_{cf_{1},..,cf_{q}}\frac{1}{q}\sum_{u=1}^{q}||\mathcal{W}(cf_{u})-y_{cf}||_{l_2} \notag\\
    s.t. \ 
    is\_proximal(cf_{u})=1,\forall u\in [1,q]\label{eq:proxfnct}\\
    is\_diverse(cf_{u})=1,\forall u\in [1,q]\label{eq:diversfnct}\\    
    is\_feasible(cf_{u})=1,\forall u\in [1,q]\label{eq:fscstrnt} %\\
\end{align}
}
\normalsize

\noindent
where $cf_{u}$ is a counterfactual explanation, $q$ is the total number of generated CFs and  $\mathcal{W}(\bullet)$ our Schedulability Boundary Function. 
The loss function $\mathcal{L}(\textbf{x}_{t_{k}})$ is also subject to the proximity, diversity and feasibility constraints (Equations \ref{eq:proxfnct}-\ref{eq:fscstrnt}).

\subsection{Feasible schedule generation}\label{ssec:trees}

{\bf Schedule Generation through Random Forests.} 
Our explainable scheduler aims to generate 
counterfactual explanations that transform infeasible task configurations into feasible ones with minimal changes,
by learning the boundary between schedulable and non-schedulable setups and efficiently exploring the solution space. 
We choose the Random Forest model, a state-of-the-art ML model for our Explainable Scheduler, as it not only facilitates boundary estimation but also enables efficient search for feasible configurations within this space. 
Random Forests (RFs) are ensemble models that combine multiple decision trees, 
making them highly accurate and less prone to overfitting. This ensures that the counterfactuals generated are based on reliable predictions.
A Random Forest trains individual trees using past data and learns 
patterns in the data that determine whether tasks are schedulable. 
They are versatile for counterfactual generation as they can work well even in the absence of pre-existing configurations. 
We propose a simple but effective approach to guide the search 
to the set of regions containing the training data points, this makes the search 
extremely fast, producing good, feasible estimates.
Furthermore, Random Forests are almost
natural to parallelize since the building of individual trees can be
distributed across multiple nodes, thereby offering near-perfect scaling. 
All these make our scheduler a powerful tool for identifying actionable configurations in a timely manner as we discuss below:

\subsubsection{{\bf Building RFs for Counterfactual Generation}}
During training, the Random Forest model learns patterns that distinguish schedulable from non-schedulable tasks, using container features such as memory, CPU cores, and the number of replicas. 
Each decision tree in the Random Forest is trained on a bootstrapped subset of the data to capture diverse patterns. Tree construction proceeds by recursively splitting nodes on randomly selected features, using thresholds such as Gini impurity or entropy to best separate schedulable from non-schedulable instances. The resulting leaf nodes are labeled as schedulable ($\mathcal{Y}=1$) or non-schedulable ($\mathcal{Y}=0$). 
Once trained, the decision trees in the Random Forest collectively identify schedulable configurations across varying features, and the final set of feasible configurations is obtained by aggregating these results.

\subsubsection{{\bf Traversing the RF for Counterfactual Generation}}\label{sssec:cftraverse} 
Given a XSchedTask instance $\textbf{x}_{t_{k}}$, the goal is to traverse the forest and obtain all configurations that make the task schedulable. 
Each decision tree in the trained Random Forest evaluates the input instance by following the splits it learned during the training phase.
At each level, we move down the tree based on the feature thresholds at each node, until we reach a leaf node where we derive a configuration ({\it i.e.,} schedulable or non-schedulable) and we select the one that satisfies the schedulability label ($\mathcal{Y}=1$) from each tree.

\subsubsection{{\bf Optimizing the traversal paths of the RF through pruning}} Traversing the entire Random Forest yields all possible counterfactual explanations, often producing an excessively large set of configurations. To reduce the search space, we introduce a pruning strategy that incorporates the proximity, diversity, sparsity, and feasibility constraints of our explainable scheduler, guiding the traversal toward acceptable regions. Prior work that has explored pruning through Lasso regularization, 
ensemble pruning, 
or parameter tuning (e.g., limiting tree depth, setting a minimum number of samples per leaf) are unsuitable in our context, as they risk discarding counterfactuals that may still provide valuable guidance to users. 
We apply pruning during inference on a given XSchedTask instance $\textbf{x}_{t_{k}}$ using an \textit{early-stopping traversal} strategy. Specifically, traversal halts once the proximity, diversity, sparsity, and feasibility constraints are satisfied and the desired label is reached, then the configuration at that node is then returned without exploring deeper branches. This approach significantly reduces inference complexity.

%% file: evaluation.tex
\section{Experimental Evaluation}\label{sec:evaluation}

\noindent
{\bf Experimental Testbed Setup and Datasets.}
We conducted experiments in our 8-node local cluster, each equipped with an Intel(R) Core(TM) i7-7820X CPU @ 3.60GHz, 128GB RAM, 128 cores, an Nvidia RTX 4090 GPU, 1Gbps Ethernet, and Ubuntu 20.04 LTS. Our workloads were derived from Alibaba’s production cluster trace \cite{alibabadata}, which records 24 hours of co-located long-running and batch jobs. Each task comprises multiple instances, and a task is marked complete only when all its instances complete; dependencies enforce that downstream tasks cannot start until all instances of upstream tasks have finished (e.g., task-2 executes after task-1).

\begin{figure}[t]
\centering
\begin{minipage}{0.45\linewidth}
\includegraphics[width=\linewidth]{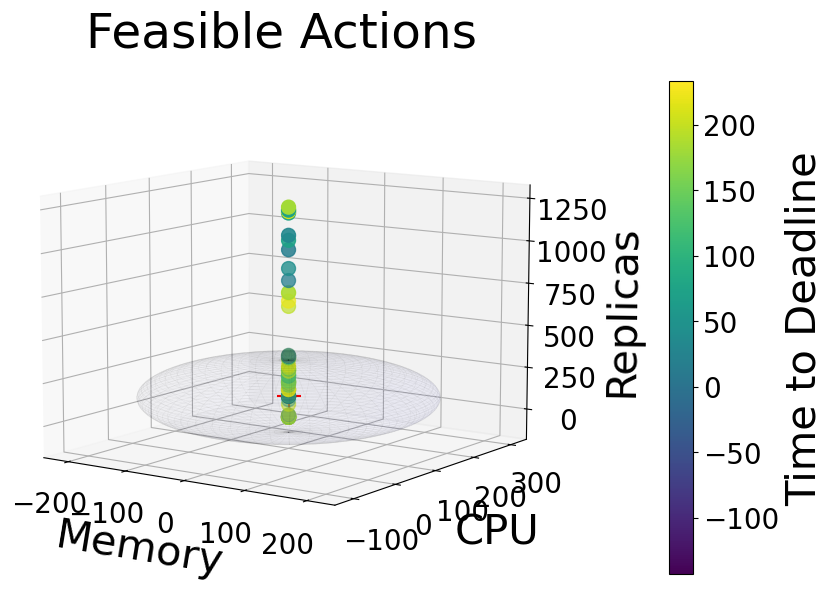}
\caption{\label{fig:feasact-plots}Feasible Actions (the gray circle denotes the area of the feasible solutions)}
\end{minipage}\hfill
\begin{minipage}{0.45\linewidth}
\includegraphics[width=\linewidth]{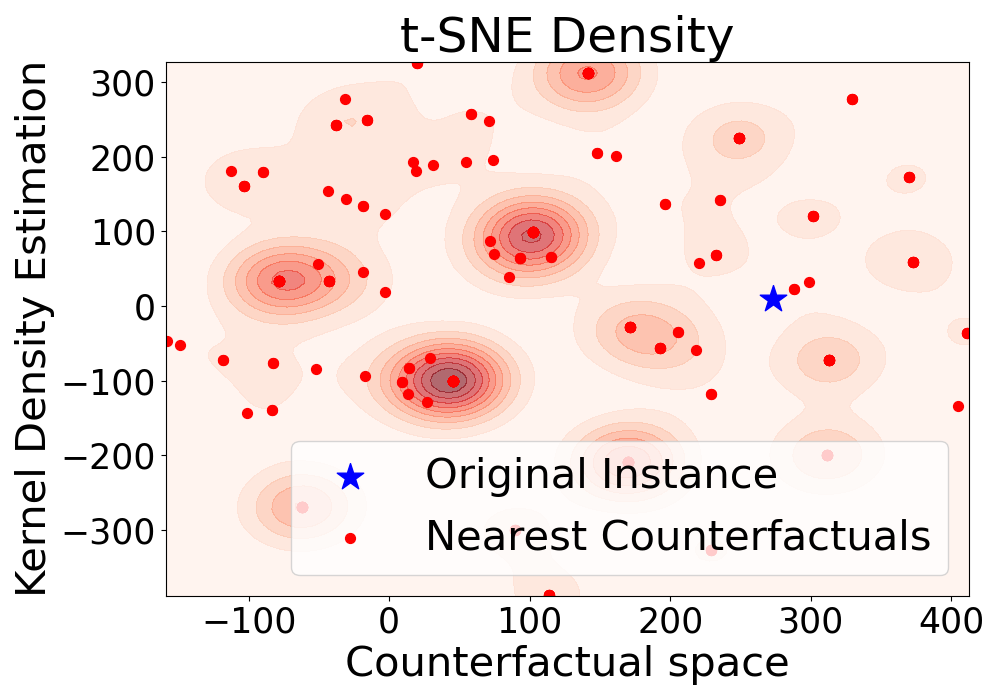}
\caption{\label{fig:tsne-plots}Density of Feasible Actions (blue star annotates the original instance)}
\end{minipage}\hfill
\end{figure}

\subsection{Evaluation}

{\bf Feasible Actions.} 
First, we evaluated the feasibility of actions generated by X-Sched to demonstrate the proximity of counterfactuals to the original instance. For a given deadline, we enumerated the possible actions and computed their time-to-deadline. Figure~\ref{fig:feasact-plots} illustrates the solution space for two user instances in a 4D plot, where the brown circle marks the requested configuration. The results highlight how container parameters (memory, CPU cores, replicas) impact deadline adherence. 
As expected, the majority of the generated counterfactual explanations concentrate in a specific area in the 3-dimensional space (drawn by a gray sphere), illustrating that the generated counterfactual explanations comply with the feasibility and the proximity constraints.

{\bf Density of feasible actions using t-SNE.} Next, we examined the concentration of counterfactual explanations to assess robustness and diversity. Using t-SNE with Kernel Density Estimation, we projected the solution space (Figure~\ref{fig:tsne-plots}). The contour plots reveal dense clusters around the original instance (drawn by the blue star), 
and their spread across regions highlights their diversity.

%% file: related.tex
\section{Related Work}

{\bf Explainability.} Explaining Machine Learning Models has been recently studied in the literature\cite{wachter2017counterfactual,DBLP:conf/fat/MothilalST20}. Counterfactual explanations were first introduced in the work of \cite{wachter2017counterfactual} and have been considered as the state-of-the-art explanation method for black-box classifiers. The work of \cite{DBLP:conf/fat/MothilalST20}, although relevant to ours, lacks key design properties of our explainable scheduler
while in \cite{icdcs2025} explainability is exploited in concert with fairness.

{\bf Scheduling.} Resource provisioning  for task completion time optimization has been the focus of many works in the literature 
\cite{DBLP:conf/bigdataconf/Zhang0WSL19,DBLP:journals/cluster/ChengHTSCL22,DBLP:conf/sc/WangW00020,hu2019spear,DBLP:conf/sosp/NarayananKAKAKB21,DBLP:conf/icdcs/CusackNGHOKRH22,zhou2019goldilocks}. The authors of \cite{DBLP:conf/bigdataconf/Zhang0WSL19} propose a DRL approach for scheduling parallel tasks in cloud environments and estimating the task completion time. Similarly, in \cite{DBLP:journals/cluster/ChengHTSCL22}, the authors propose a cost-aware DRL technique for scheduling real-time job requests, while in \cite{DBLP:conf/sc/WangW00020} they propose a general-purpose scheduler that learns to optimally place LRA containers using DRL. 
The work of \cite{hu2019spear} proposes an optimized DRL-based dependency-aware task scheduling approach that minimizes the makespan of complex jobs. In \cite{DBLP:conf/sosp/NarayananKAKAKB21} they solve the resource allocation problem by partitioning it into subproblems. 
Escra\cite{DBLP:conf/icdcs/CusackNGHOKRH22} enables event-based resource allocation for a single container and distributed resource allocation to manage a collection of containers. Finally, in \cite{zhou2019goldilocks} they propose a graph partitioning algorithm for developing a resource provisioning system. In contrast, X-Sched offers an explainable scheduler that not only enables tasks to meet their deadlines but also guides users to choose effective configurations.

%% file: conclusion.tex
\section{Conclusions}
We presented X-Sched, an explainable middleware for the scheduling of big data tasks in containerized environments. X-Sched uses explainability techniques to generate
actionable guidance on resource configurations that makes task
execution in containerized environments both feasible and practical, under resource
and time constraints.